\documentclass[onecolumn,showpacs]{revtex4}

\usepackage{natbib}
\usepackage{epstopdf}
\usepackage{graphicx}
\usepackage{dcolumn}

\usepackage{eucal}
\usepackage[dvips]{epsfig}
\usepackage{amssymb}
\usepackage{amsmath,amsthm}

\newcommand{\be}{\begin{eqnarray}}
\newcommand{\ee}{\end{eqnarray}}

\newcommand{\bse}{\begin{subequations}}
\newcommand{\ese}{\end{subequations}}

\newcommand{\bnum}{\begin{enumerate}}
\newcommand{\enum}{\end{enumerate}}

\newcommand{\bit}{\begin{itemize}}
\newcommand{\eit}{\end{itemize}}

\newcommand{\bc}{\begin{cases}}
\newcommand{\ec}{\end{cases}}

\newcommand{\bpm}{\begin{pmatrix}}
\newcommand{\epm}{\end{pmatrix}}

\newcommand{\bvm}{\begin{vmatrix}}
\newcommand{\evm}{\end{vmatrix}}

\newcommand{\bs}{\boldsymbol}

\newcommand{\mrm}{\mathrm}

\newcommand{\ga}{\alpha}

\newcommand{\gc}{\gamma}
\newcommand{\gd}{\delta}

\newcommand{\gl}{\lambda}
\newcommand{\gk}{\kappa}
\newcommand{\go}{\omega}
\newcommand{\gt}{\theta}

\newcommand{\Gc}{\Gamma}

\newcommand{\Gd}{\Delta}

\newcommand{\f}{\frac}
\newcommand{\diff}{\mrm{d}}

\newcommand{\lan}{\langle}
\newcommand{\ran}{\rangle}

\newcommand{\kBT}{k_B\Temp}

\newcommand{\csp}{\;,\qquad}

\newcommand{\dt}{\diff t}

\newcommand{\met}{\mrm{m}}

\renewcommand{\sec}{\mrm{s}}
\newcommand{\kg}{\mrm{kg}}

\renewcommand{\kBT}{kT}

\begin{document}

\title{Swimmer-tracer scattering at low Reynolds number}

\author{J\"orn Dunkel}
\email{jorn.dunkel@physics.ox.ac.uk}
\author{Victor B. Putz}
\author{Irwin M. Zaid}
\author{Julia M. Yeomans}
\affiliation{Rudolf Peierls Centre for Theoretical Physics, University of Oxford, 1 Keble Road, Oxford OX1 3NP, United Kingdom}
\date{\today}

\begin{abstract}
Understanding the stochastic dynamics of tracer particles in active fluids is important for identifying the physical properties of flow generating objects such as colloids, bacteria or algae. Here, we study both analytically and numerically the scattering of a tracer particle in different types of time-dependent, hydrodynamic flow fields. Specifically, we compare the tracer motion induced by an externally driven colloid with the one generated by various self-motile, multi-sphere swimmers. Our results suggest that force-free swimmers generically induce loop-shaped tracer trajectories.  The specific topological structure of these loops is determined by the hydrodynamic properties of the microswimmer. Quantitative estimates for typical experimental conditions imply that the loops survive on average even if Brownian motion effects are taken into account. \end{abstract}

\maketitle

\section{Introduction}
Scattering processes are ubiquitous in Nature, ranging from  elementary particle collisions at subatomic scales to gravitational encounters in galactic clusters. Traditionally,   scattering experiments have played an important role in elucidating the interactions between \lq non-living\rq\space physical objects. Nowadays, modern experimental techniques allow us to track the motion of individual microorganisms~\cite{2000Be,2007SoEtAl,2009CoWe,2009Gi}, as beautifully illustrated by recent high-speed microscopy observations of \textit{Volvox}~\cite{2009DrEtAl_Gold} and  \textit{Chlamydomonas reinhardtii}~\cite{2009LeEtAl_Gold,2009PoEtAl_Gold}. These and similar experiments on colloidal systems~\cite{2009LeEtAl} suggest that it should be possible  in the near future to systematically study the effective hydrodynamic interaction forces generated by algae, bacteria or artificial microswimmers~\cite{2005DrEtAl,2008TiEtal}  through suitably designed biophysical scattering experiments.
\par
Additional motivation for studying swimmer-tracer scattering comes from recent  experiments by Leptos et al.~\cite{2009LeEtAl_Gold}, who observed that a passive tracer particle exhibits non-Gaussian diffusion when surrounded by a dilute suspension of self-motile~\emph{Chlamydomonas reinhardtii} algae. Qualitatively, the anomalous diffusive behavior arises because the algae modify the velocity field of the fluid, thereby occasionally accelerating the tracer particle to relatively large velocities.  A satisfactory quantitative explanation for this phenomenon is still missing, and a better understanding of the underlying elementary scattering processes is a key step en route to deriving the  corresponding generalized diffusion equation.
\par
The present paper  aims to identify generic features that should be observable when a small colloidal tracer particle is scattered by an artificial or natural microswimmer. Specifically, we address the following problems: How does  the tracer dynamics differ depending on whether a swimmer's motion is externally forced or internally  generated? Is it sufficient to consider an effective time-averaged description in order to correctly predict the tracer trajectories -- or is it necessary to use a time-resolved model that captures the details of the swimming stroke? How does Brownian motion, which plays a non-neglible role for (sub-)micron-sized colloidal particles, affect the tracer dynamics? To answer these questions, we will compare both analytically and numerically the dynamics of a tracer in the presence of various simplified model swimmers. 
\par
The first part of our discussion focusses on deterministic scattering processes. In order to clarify how the effective interaction range determines the asymptotic state of the tracer, we shall consider a simplified power-law interaction model and, as a more realistic example, tracer displacement due to an externally driven  colloid. The main part of the paper is dedicated to the scattering of a tracer by 2-sphere and 3-sphere swimmers that generate predominantly dipolar and quadrupolar flow fields, respectively.  We show that, asymptotically, the tracer motion in the presence of a self-motile, force-free swimmer converges to a closed loop.  The shape and direction of the loops is a signature of the specific properties of a given swimmer.  For experimentally accessible parameters, Brownian motion effects may dominate the trajectories of an individual tracer particle. In the last part of the paper, we will demonstrate that, after averaging over a few hundred to thousand samples, the mean tracer trajectories look very similar to those obtained in the deterministic limit.

\section{Mathematical model}
We consider the motion $\bs x(t)$ of a passive, colloidal tracer particle in a fluid due to the presence of an active object (\lq swimmer\rq). The latter, which may be an externally forced colloid,  an artificial microswimmer or a micororganism (e.g.,  an  alga or a bacterium), is described by a phase space coordinate vector~$\Gc(t)$. For example, for a spherical colloid with position vector $\bs X(t)$ and velocity  $\dot{\bs X}(t):=\diff\bs X(t)/\dt$, $\Gc(t)$ is simply given by $\Gc(t)=(\bs X(t),\dot{\bs X}(t))$. For more complex objects, such as multi-sphere swimmers~\cite{2007PoAlYe,2008GoAj,2009AlPoYe}, $\Gc$ comprises all the coordinates necessary to uniquely specify the motion of the swimmer. Throughout, we shall assume that, in good approximation,  the tracer particle does not affect the swimmer motion, i.e.,  $\Gc(t)$ is taken to be independent of $\bs x(t)$.  

\subsection{Langevin dynamics of the tracer}
In principle, when studying the scattering of a tracer particle by a self-swimming object, one can distinguish (at least) two different approaches:
\begin{itemize}
\item[(i)] analyzing the approximate mean motion of the tracer particle in the time-averaged (or stroke-averaged) effective flow field of the swimmer.  
\item[(ii)]  computing the exact, time-resolved motion of the tracer particle in the full oscillatory flow field of the swimmer.  
\end{itemize}
Below we will compare results from both methods for  multi-sphere swimmer models~\cite{2008GoAj,2009AlPoYe}. We will start from the time-averaged description (i) which is less accurate, but allows us to obtain analytical estimates. When adopting this coarse-grained approach,  the swimmer state is approximated by $\Gc(t)=(\bs X(t),\dot{\bs X}(t))$, and the swimmer is assumed to move at constant velocity $\dot{\bs X}(t)\equiv \bs V$ along the straight line  
\be
\bs X(t)=\bs X_0+ t \bs V.
\ee
The effective stroke-averaged swimmer velocity $\bs V$ results from the microscopic swimmer dynamics. 
\par
Let us assume we know the velocity field $\bs v(\bs x| \Gc(t))$ that is generated by a given swimmer configuration $\Gc(t)$. In this case, we can model the motion of a micron-sized tracer particle by the overdamped  Langevin equation 
\bse
\be\label{e:LE-a}
\dot{\bs x}(t) = \bs v(\bs x(t)| \Gc(t))+ (2D_0)^{1/2}\bs \xi(t).
\ee
This equation is valid in the Stokes (zero Reynolds number) regime, which is realized to good approximation under typical  experimental conditions~\cite{2009DrEtAl_Gold,2009LeEtAl_Gold,2009PoEtAl_Gold}. The second contribution on the rhs of  Eq.~\eqref{e:LE-a} is Gaussian white noise $\bs \xi(t)=(\xi_i(t))$, which describes thermal fluctuations in the fluid and is  characterized by 
\be
\lan \xi_i(t)\ran=0\csp
\lan \xi_i(t)\xi_j(s)\ran=\gd_{ij}\gd(t-s).
\ee
For a spherical tracer particle of radius $a_0$, the thermal diffusion constant $D_0$ is given by the Stokes formula
\be
D_0={\kBT}/{\gc_0}={\kBT}/({6\pi\eta a_0}),
\ee
\ese
where $a_0$ is the tracer radius and $\gc_0$ the Stokes friction coefficient. For example,  considering  $a_0=1\mu\met$ and water at room temperature $T=25^\circ$C with viscosity $\eta=10^{-3} \kg\,\met^{-1}\sec^{-1}$ and $\kBT=4.11\times 10^{-21}\kg\,\met^2\sec^{-2}$, we have $D_0= 0.22\mu\met^2\sec^{-1}$.
\begin{figure}[h]
\centering
\includegraphics[width=7cm]{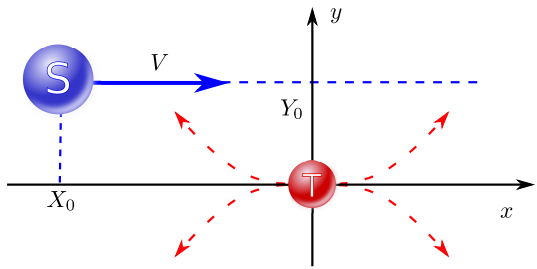}
\caption{\emph{Swimmer-tracer scattering.} The swimmer (S) starts at $\bs X_0=(X_0,Y_0,0)$  and moves with velocity $\bs V=(V,0,0)$  parallel to the $x$-axis. The trajectory of the tracer  (T), initially located at $\bs x(0)=(0,0,0)$, depends on the flow field generated by the swimmer. \label{fig01_Scatter}}
\end{figure}

\subsection{Initial conditions}
We intend to analyze swimmer-tracer scattering processes for well-defined, reproducible initial conditions.  For this purpose, we choose the coordinate system such that the spherical tracer particle is initially located at the origin, $\bs x(0)=\bs 0$. The swimmer (characteristic length scale $A$) starts at $\bs X_0=(X_0,Y_0,0)$ with $|Y_0|>a_0+A$ and moves in the $(z=0)$-plane  with  an average velocity $\bs V=(V,0,0)$  parallel to the $x$-axis, see Fig.~\ref{fig01_Scatter}. 

\section{Tracer scattering in time-dependent flow fields} 
\label{s:swimmer_velocity_fields}
Our main objective is to understand the motion of the tracer in the velocity field  $\bs v(\bs x| \Gc(t))$ of an externally driven or  self-motile object.  In order to simplify the notation, it is convenient to define the distance vector  between tracer and swimmer by 
\bse
\be
\bs r(t)=\bs x(t)-\bs X(t),
\ee
and to introduce two unit vectors
\be\label{e:unit_vectors}
\hat{\bs r}(t)=\f{\bs r(t)}{|\bs r(t)|} \csp
\hat{\bs n}=\f{\bs V}{|\bs V|},
\ee
\ese
with $\hat{\bs n}(t)$ characterizing the orientation of the swimmer motion. In the examples considered below, the effective hydrodynamic interaction  between swimmer and tracer is mediated by non-diagonal tensors $\bs H=\bs H(|\bs r|,\hat{\bs r},\hat{\bs n})=(H_{ij})$, which relate the fluid velocity at the position $\bs x(t)$ to the velocity $\bs V$ of the swimmer through
\be
\bs v=\bs H \bs V.
\ee 
\par
The exact structure of $\bs H$ encodes  the details of the propulsion mechanism. For example,  consider an externally forced colloid of radius $A$ that is pulled at constant velocity  $\bs V$. If the colloid passes the tracer particle at a sufficiently large distance $|\bs r(t)|:=|\bs x(t)-\bs X(t)|\gg A +a_0$, then $\bs H$ is given by the Oseen tensor
\be\label{e:Oseen}
H_{ij}=\f{3A}{4|\bs r|}\left(\gd_{ij}+\hat{r}_i\hat{r}_j\right).
\ee
By contrast, the effective velocity field generated by a self-swimming microorganism usually decays more rapidly as  $|\bs r|^{-\ga}$, $\ga\ge 2$. In Sections~\ref{s:colloid}, \ref{s:dipolar} and~\ref{s:quadrupolar}  we shall compare the scattering of tracers by externally forced colloids and multi-sphere swimmers. However, in order to demonstrate how the effective interaction range $\ga$ affects  the scattering process,  it is instructive to analyze a simplified power law interaction model first.

\subsection{Simplified power law  interaction ($\ga$-model)}
\label{s:alpha}
Let us assume that a hypothetical \lq swimmer\rq~of radius $A$ generates a fluid velocity field of the simple power law form
\be\label{e:alpha}
\bs v(\bs x| \bs X(t),\bs V)
=
\bs V\gk\; \left(\f{A}{| \bs r(t)|}\right)^\ga.
\ee
The dimensionless parameter~$\gk$ quantifies the interaction strength. In order to understand the basic dynamics of a tracer in this field, we will focus on the deterministic limit $D_0=0$ in the Langevin equation~\eqref{e:LE-a}. Brownian motion effects (corresponding $D_0>0$) will be discussed later  for  more realistic flow fields.
\par
\begin{figure}[t]
\centering
\includegraphics[width=8.5cm]{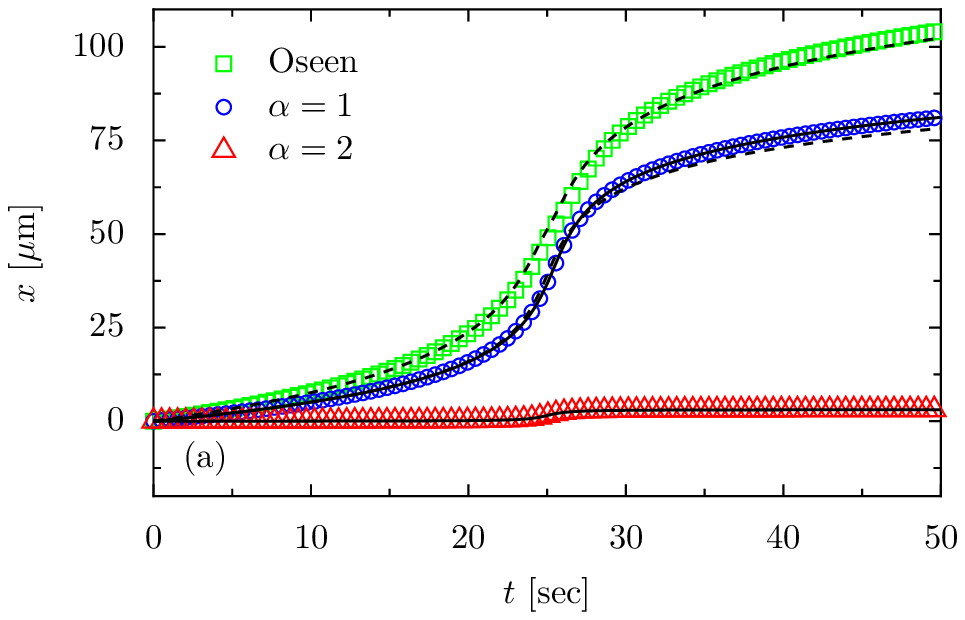}
\includegraphics[width=8.5cm]{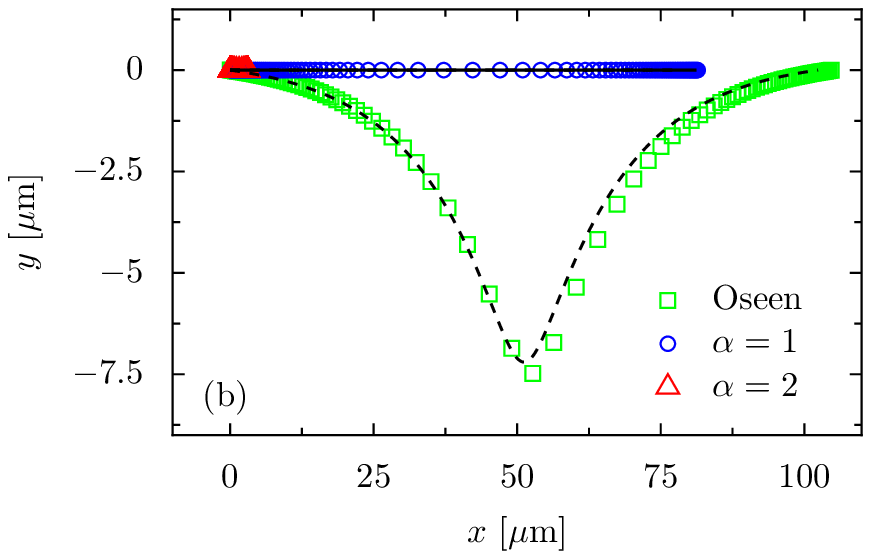}
\caption{
\emph{Deterministic tracer trajectories for three different model interactions.} (a) Tracer displacement in the direction of the swimmer's motion as a function of time~$t$ for the simplified $\ga$-model with $\gk=1$ and the Oseen interaction. (b) Tracer trajectories in the $(x,y)$-plane. For colloidal-type long-range interactions that scale with distance as $|\bs r|^{-1}$ the tracer is dragged to infinity  - the divergence is logarithmic in time. By contrast,  for shorter range interactions with $|\bs r|^{-\ga},\ga\ge 2$ the maximal tracer displacement remains finite, see red curve/triangles.  Solid lines correspond to the exact result~\eqref{e:toy_exact}, and dashed lines to the approximate far-field result~\eqref{e:scatter_simple_1}. 
The symbols were obtained by numerically integrating the deterministic equations of motion~\eqref {e:LE-a} with $D_0=0$ using a simulation time step $\Gd t=2.2\times 10^{-7}\sec$. Parameters: swimmer velocity $V=100\mu/\sec$, swimmer radius $A=10\mu \met$,  tracer radius $a=1\mu\met$,  impact parameter $Y_0=10A$, initial swimmer position $X_0=-25A$.
\label{fig02_toy}
}
\end{figure}

Adopting the  initial conditions   $\bs V=(V,0,0)$, $\bs X(0)=(X_0,Y_0,0)$,  and $\bs x(0)=\bs 0$ as depicted in Fig.~\ref{fig01_Scatter}, the swimmer moves in the $x$-direction and its $x$-position at time $t$ is given by $X(t)=X_0+tV$. The isotropic power law model~\eqref{e:alpha} corresponds to the diagonal tensor \mbox{$H_{ij}=\gk\,(A/|\bs r|)^\ga \gd_{ij}$}, which implies that  the tracer particle also moves along the $x$-axis, i.e., $\bs x(t)=(x(t),0,0)$ where
\be\label{e:toy_eom}
\dot x(t) =V\gk \left(\f{A}{\{[X(t) - x(t)]^2+Y_0^2\}^{1/2}}\right)^\ga. 
\ee
It is possible to find exact, implicit solutions of Eq.~\eqref{e:toy_eom} for $\ga=1,2,\ldots$ by transforming to the comoving swimmer frame. For example, for long-range interactions with $\ga=1$, we obtain, in the original coordinate system, 
\bse\label{e:toy_exact}
\be
x(t) 
=\notag
\gk A \biggl\{\log\left(\frac{X(t)-x(t)+\sqrt{[X(t)-x(t)]^2+Y_0^2}}{X_0+\sqrt{X_0^2+Y_0^2}}\right) -
\frac{\gk A }{\sqrt{Y_0^2-\gk^2 A ^2}}\biggl[
\arctan\left(\frac{\gk A X_0  }{\sqrt{X_0^2+Y_0^2} \sqrt{Y_0^2-\gk^2 A ^2}}\right) -
\\
\arctan\left( \frac{ \gk A[X(t)-x(t)]}{\sqrt{[X(t)-x(t)]^2+Y_0^2} \sqrt{Y_0^2-\gk^2 A ^2}}\right)+
\arctan\left(\frac{X_0}{\sqrt{Y_0^2-\gk^2 A ^2}}\right)-
\arctan\left(\frac{X(t)-x(t)}{\sqrt{Y_0^2-\gk^2 A ^2}}\right)
\biggr]\biggr\},
\quad
\ee
whereas for shorter range interactions with $\ga=2$
\be
x(t)=\frac{\gk A^2}{\sqrt{Y_0^2-\gk A^2 }}
\biggl[ \arctan\left(\frac{X(t)-x(t)}{\sqrt{Y_0^2-\gk A^2}}\right)
-\arctan\left(\frac{X_0}{\sqrt{Y_0^2-\gk A^2 }}\right) 
 \biggr].
\ee
\ese
The exact results~\eqref{e:toy_exact} are transcendental equations for $x(t)$ which can be numerically inverted for  a given value $t>0$ to give the trajectory of the tracer particle. In particular, these equations cover both the near- and far-field scattering behavior. However, to gain additional analytic insights, it is useful to consider in more detail the far-field scattering, corresponding to  $|x(t)|\ll \max \{|Y_0|,|X(t)|\}$  and $\gk^{3-\ga} A^2\ll Y_0^2$. In this limit one obtains, for $\ga=1$, the approximate result
\bse\label{e:scatter_simple_1}
\be\label{e:scatter_simple_1a}
x(t)\simeq
\gk A  \log\biggl[\f{X(t)+\sqrt{X(t)^2+Y_0^2}}{X_0+\sqrt{X_0^2+Y_0^2}}\biggr].
\ee
Inserting $X(t)=X_0+tV$, we see that asymptotically  (for $t\to \infty$)
\be
x(t)\simeq
\gk A   \log\left[\f{2t V}{X_0+\sqrt{X_0^2+Y_0^2}}\right].
\ee
\ese
Thus, for $\bs v \propto |\bs r|^{-1}$ the solution is logarithmically divergent in time, implying that  the tracer particle is slowly dragged away to infinity. Defining the asymptotic displacement  as a function of $\ga$ by 
\be
(\Gd x)^\infty_\ga:= \lim_{t\to\infty}[x(t)-x(0)],
\ee
the divergence implies $(\Gd x)^\infty_1=\infty$. Performing a similar analysis for $\ga=2$ yields
\bse\label{e:scatter_simple_2}
\be\label{e:scatter_simple_2a}
x(t)\simeq
\gk A \left(\f{A}{|Y_0|}\right)  \left[\arctan\left(\frac{X(t)}{|Y_0|}\right)-\arctan\left(\frac{X_0}{|Y_0|}\right)\right],
\qquad
\ee
and thus, in the limit $t\to \infty$,
\be\label{e:scatter_simple_2b}
(\Gd x)^\infty_2=
\gk A \biggl(\f{A}{|Y_0|}\biggr)
 \left[\f{\pi}{2}-\arctan\left(\frac{X_0}{|Y_0|}\right)\right],
\ee
\ese
i.e., for $\ga=2$  the tracer particle comes to rest after a finite distance $(\Gd x)^\infty_2$. Analogous results can be obtained for arbitrary values $\ga\ge 2$. For example, for $\ga=3$ 
\bse
\be
x(t)\simeq
\gk A \left(\f{A}{|Y_0|}\right)^2  \left[
\frac{X(t)}{\sqrt{X(t)^2+Y_0^2}}-\frac{X_0}{\sqrt{X_0^2+Y_0^2}}
\right],
\ee
yielding the maximal displacement
\be
(\Gd x)^\infty_3 =
\gk A\; \biggl(\f{A}{|Y_0|}\biggr)^2 \left(
1-\frac{X_0}{\sqrt{X_0^2+Y_0^2}}
\right).
\ee
\ese
To briefly summarize, for long-range hydrodynamic interactions with $\bs v \propto |\bs r|^{-1}$  the tracer experiences an infinite displacement, whereas for shorter range interactions with $\bs v\propto |\bs r|^{-\ga},\ga \ge 2$ the displacement is finite. These results are illustrated graphically in Fig.~\ref{fig02_toy}.
\par
The simple power law model~\eqref{e:alpha} is useful for illustrating how the range of the effective hydrodynamic interactions affects the asymptotic state of the tracer particle. However, this truncated model neglects fluid velocity components that act perpendicular to the swimmer's direction of motion. In the remainder, we will focus on more physical models that account for these effects. As the first example, we consider the scattering of a tracer by a driven colloid.

\subsection{Externally forced colloid}
\label{s:colloid}
Consider a spherical colloid (radius $A$)  dragged at constant velocity $\bs V$ past a smaller tracer particle (e.g., by a movable laser trap). If the distance between tracer and colloid is sufficiently large, $|\bs r(t)|\gg A+a_0$, the velocity field components $v_i=H_{ij} V_j$  experienced by the tracer are determined by the Oseen tensor~\eqref{e:Oseen} and we can approximate $\bs r(t)\simeq \bs X(t)$. Considering the deterministic limit  $D_0=0$ and the same initial conditions as before, we then find that the trajectory of the tracer particle in the $(z=0)$-plane is determined by the differential equation
\be\label{e:Oseen_eom}
\begin{pmatrix}
\dot x\\
\dot y
\end{pmatrix}
=
\gk V \f{A }{\sqrt{X(t)^2+Y_0^2}}
\biggl[\!
\begin{pmatrix}
1\\
0
\end{pmatrix} +
\f{X(t)}{X(t)^2+Y_0^2} 
\begin{pmatrix}
X(t)\\
Y_0
\end{pmatrix} 
\biggr].\;\;\;
\ee
Owing to the non-vanishing off-diagonal components of the Oseen tensor, the tracer experiences a transverse force (velocity field) component in the $y$-direction.  Equation~\eqref{e:Oseen_eom} can be directly integrated, yielding
\bse\label{e:Oseen_result}
\be
x(t)
&=&
\gk A    
\biggl\{2 \log\left[\f{X(t)+\sqrt{X(t)^2+Y_0^2}}{X_0+\sqrt{X_0^2+Y_0^2}}\right]+
\frac{X_0}{\sqrt{X_0^2+Y_0^2}}-\frac{X(t)}{\sqrt{X(t)^2+Y_0^2}}\biggr\},
\label{e:Oseen_result_a}
\\
y(t)
&=&
\gk A    \left[\frac{Y_0}{\sqrt{X_0^2+Y_0^2}}-\frac{Y_0}{\sqrt{X(t)^2+Y_0^2}}\right],
\ee
\ese
where $X(t)=X_0+tV$. Letting $t\to\infty$, we find that asymptotically
\bse
\be\label{e:Oseen_result_asymp_a}
x(t)
&\simeq&
2\gk A    \log\left[\f{2t  V}{X_0+\sqrt{X_0^2+Y_0^2}}\right],\\
y(t)
&\to&
\gk A   \frac{Y_0}{\sqrt{X_0^2+Y_0^2}}.
\ee
\ese
Note that the logarithmic divergence predicted by Eq.~\eqref{e:Oseen_result_asymp_a} is in agreement with the earlier result~\eqref{e:scatter_simple_2a}, i.e. the tracer particle slowly follows the colloid to infinity.
\par
The deterministic result~\eqref{e:Oseen_result} can be expected to correctly describe a single tracer trajectory only if Brownian fluctuations are negligible corresponding to the low-temperature, high-viscosity regime.  For a micron-sized tracer particle  in water at room temperature,  Brownian motion effects can become relevant if the speed $V$  of the \lq swimmer\rq~colloid is too  low and/or the distance between tracer and swimmer colloid becomes too large. In this case, Eq.~\eqref{e:Oseen_result} provides an approximate description of the ensemble mean when the scattering experiment is repeated using identical (i.e., deterministic) initial conditions. The statistical fluctuations around the mean trajectory~\eqref{e:Oseen_result} can be estimated by $\sqrt{2D_0 t}$, where $D_0\simeq 0.22\mu\met^2/\sec$ for a micron-sized tracer. This suggests that for the parameters and initial conditions used in Fig.~\ref{fig02_toy}, the characteristic features of the solution~\eqref{e:Oseen_result} should remain observable even for a single tracer trajectory.
\vspace{0.3cm}
\par
Having discussed the scattering of a tracer by an externally driven colloid, we shall focus on self-motile swimmers
in the remainder of the paper~\cite{2008AlYe,2008LaBa,2008GoAj,2009AlPoYe,2009DuZa,2010PuDuYe}. In particular, we would like to identify characteristic features of the tracer trajectories that provide a signature of truly self-swimming objects, which do not generate a net force during a swimming stroke. Specifically, we are interested in understanding how the tracer trajectories differ depending on
\begin{itemize}
\item[(i)] whether a force-free swimmer  is  extensile (\lq pusher\rq) or contractile (\lq puller\rq) ;
\item[(ii)] whether a force-free swimmer generates a dipolar or a quadrupolar flow field. 
\end{itemize}
For this purpose, we will study simple multi-sphere swimmers that are analytically tractable and can be easily tuned from extensile to contractile. An additional advantage of the simplified models is that the underlying  microscopic swimming mechanisms can be implemented numerically using the methods described in Ref.~\cite{2009DuZa,2010PuDuYe}. This allows us to compare analytic estimates, which are based on a time-averaged description, with numerical results for the  exact time-resolved dynamics.  

\begin{figure}[t]
\centering
\includegraphics[width=7cm]{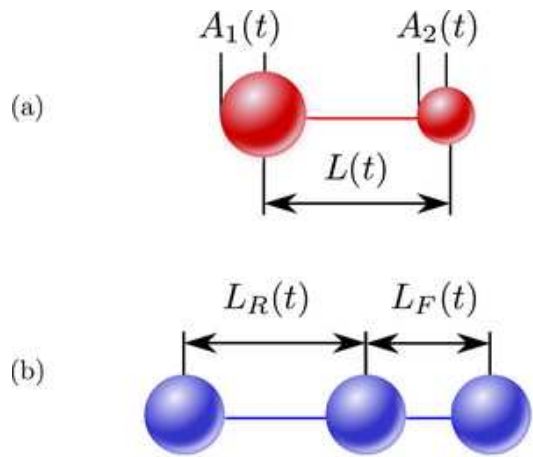}
\caption{
\label{fig03_swimmer}
\emph{Simple multi-sphere swimmer models.} (a) The 2-sphere swimmer with variable radius spheres  generates a dipolar net flow field that can be either extensile or contractile with respect to the swimmer axis, see Figs.~\ref{fig0x_inch_flow_averaged}. (b) The 3-sphere swimmer~\cite{2008GoAj} generates a flow field that is essentially quadrupolar~\cite{2009AlPoYe}, see  Figs.~\ref{fig0x_gol_flow}. Consequently, the trajectories of tracer particles are qualitatively different when scattered by the two swimmers.
}
\end{figure}

\subsection{Dipolar two-sphere swimmer}
\label{s:dipolar}
As the first example of tracer scattering by a self-motile object, we consider a slender 2-sphere swimmer consisting of two beads (radii $A_{1,2}$)  at positions $\bs X_{1,2}(t)$  that are connected by an oscillating leg of length 
\be\label{e:2-length}
L(t)=\ell+\xi \sin(\go t+ \gt)
\csp \ell \gg \xi>0,
\ee
see Fig.~\ref{fig03_swimmer} (a). The orientation vector of the swimmer is defined by
\be
\hat{\bs n}(t):=\f{ \bs X_2(t)-\bs X_1(t) }{ | \bs X_2(t)-\bs X_1(t) | }.
\ee
If the sphere radii $A_{1}$ and $A_2$ are constant then this 2-sphere swimmer reduces to an oscillating dumbbell~\cite{2008AlYe,2008LaBa,2010PuDuYe}, which is prevented from swimming by the scallop theorem~\cite{1977Pu,2008LaBa}. A simple, self-motile 2-sphere swimmer can be obtained if the size of the spheres changes periodically in time  according to 
\be
A_1(t)=A+\gl\sin(\go t+\phi_1)
\csp
A_2(t)=A+\gl\sin(\go t+\phi_2)
\csp
A\gg \gl>0.
\ee
For simplicity, we neglect secondary flow contributions produced by the expansion and shrinkage of the spheres by assuming that (to good approximation) they merely change their hydrodynamic resistance in a manner consistent with a changed radius.  Considering the limit $\gl,\xi \ll A,\ell$ and averaging over a stroke-period $[t,t+2\pi/\go]$, one can show that  the 2-swimmer moves at an average velocity $\bs V=V\hat{\bs n}$, where
\bse\label{e:dumb_vel}
\be
V=\f{1}{4}\go\gl\left( \f{ \xi}{A}\right)\left(1-\f{3A}{2\ell}\right)^{-1}\sin(\Gd\phi)  \cos(\Gd \gt)
\ee
and
\be
\Gd\phi:=\f{1}{2}(\phi_1-\phi_2)
\csp
\Gd\gt:=\f{1}{2}(\phi_1+\phi_2)-\gt.
\ee
\ese
The derivation of Eq.~\eqref{e:dumb_vel} is analogous to those for other multi-sphere swimmers~\cite{2004NaGo,2009AlPoYe}. By choosing  the phase parameters $\phi_1,\phi_2$, and $\gt$ such that $\Gd\phi\in(0,\pi)$ and $\Gd\gt\in(-\pi/2,\pi/2)$, we have $V>0$, i.e., the swimmer moves in the direction of positive $\hat{\bs n}$.
\par
\begin{figure}[t!]
\centering
\includegraphics[width=7cm]{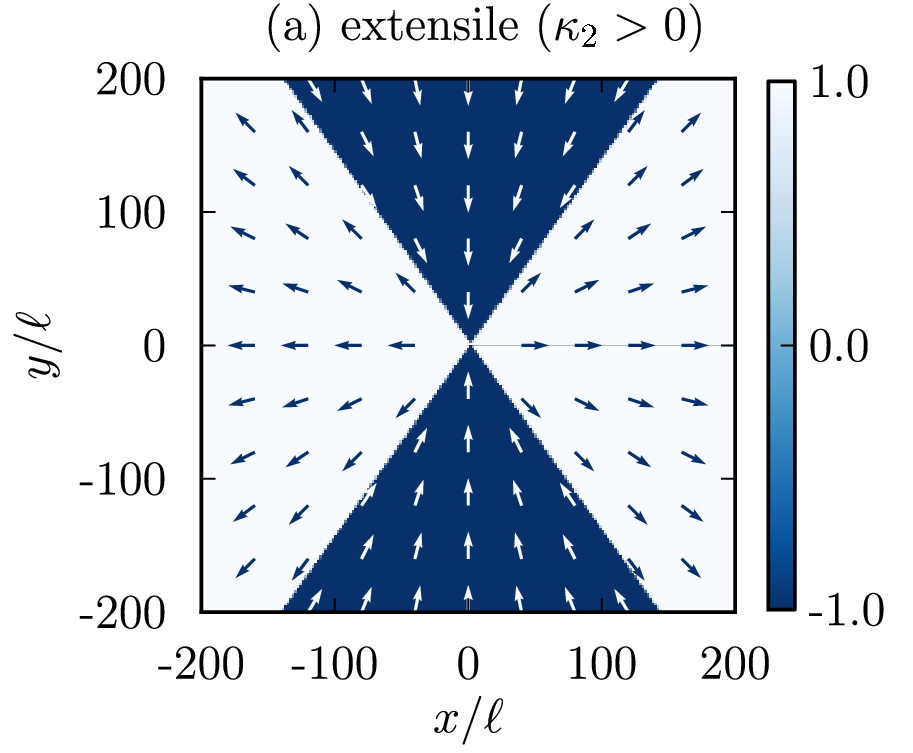}
\hspace{0.5cm}
\includegraphics[width=7cm]{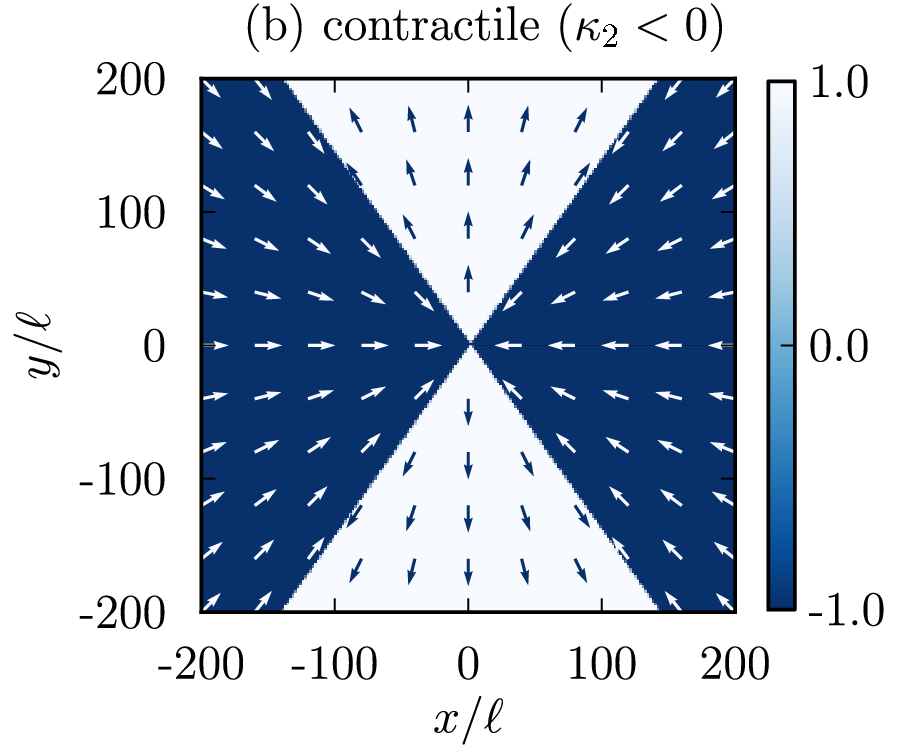}
\caption{
\label{fig0x_inch_flow_averaged}
\emph{Dipolar fluid flows generated by 2-sphere swimmer.--} The diagrams show  the stroke-averaged fluid velocity field~\eqref{e:dumb_flow}  for two different swimmer setups: (a)~extensile pusher with $\gk_2>0$ and (b)~contractile puller with $\gk_2<0$.  In both cases the swimmer is located at the origin, points along the $x$-axis, and swims in the positive $x$-direction. Swimmer parameters: $\ell=20\mu\met, \xi= A=0.1\ell, \gl=0.1A,  \go=5\times10^2\sec^{-1}$ with $\Gd\phi=\Gd\gt=\pi/4$ for extensile swimmers and $\Gd\phi=3\pi/4,\,\Gd\gt=\pi/4$ in the contractile case, yielding $V\simeq 14.7\mu\met/\sec$ and $\gk_2=\pm 8.8$. The flow fields are rotationally invariant about the $x$-axis. Arrows indicate the local direction $\hat{\bs v}:=\bs v/|\bs v|$ of the stroke-averaged flow~\eqref{e:dumb_flow}. Dark (bright) areas show the projection  $(\hat{\bs r}\hat{\bs v})$,  corresponding to  regions where the mean flow points towards (away from) the swimmer. 
}
\end{figure}

The stroke-averaged velocity field generated by the swimmer at large distances is obtained by summing the Stokeslets of the two spheres, performing a far-field Taylor (multipole) expansion and averaging the leading far-field contributions over a swimming stroke~\cite{2009AlPoYe}. By means of this procedure, one finds that the averaged far-field fluid flow generated by the swimmer is \emph{dipolar}, 
\bse\label{e:dumb_flow}
\be
\bs v(\bs x(t)| \bs X(t),\bs V)
&=&
V\gk_2 \left(\f{ A}{|\bs r(t)|}\right)^2
\left[   
3 (\hat{\bs n} \hat{\bs r})^2-1     
\right]  \hat{\bs r},
\ee
with unit vectors $\hat{\bs n}$ and $\hat{\bs r}$ as defined in Eq.~\eqref{e:unit_vectors} and
\be
\gk_2=\f{3}{4}\left(\f{\ell}{A}\right) \left(1-\f{3A}{2\ell}\right)^{-1}\f{\tan(\Gd\gt)}{\tan(\Gd\phi)}.
\ee
\ese
Note that, depending on the choice of $\Gd\phi$ and $\Gd\gt$, the coefficient $\gk_2$ can be either negative or positive for a swimmer with $V>0$. Swimmers with  $\gk_2>0$ are extensile \lq pushers\rq, whereas negative values $\gk_2<0$  correspond to contractile \lq pullers\rq. In our simulations, we used $\gt=0$, $\Gd\phi=\Gd\gt=\pi/4$ to realize an extensile swimmer and $\gt=0$, $\Gd\phi=3\pi/4,\,\Gd\gt=\pi/4$ in the contractile case.  The structure of the dipolar flow fields~\eqref{e:dumb_flow}  for these parameter values is illustrated in Figs.~\ref{fig0x_inch_flow_averaged}. The arrows show  the normalized velocity field  $\hat{\bs v}:=\bs v/|\bs v|$.  The color shading indicates the projection $(\hat{\bs r}\hat{\bs v})$, i.e., dark (bright) areas correspond to regions where the mean flow points towards (away from) the swimmer.

\paragraph*{Tracer motion in the deterministic limit.--}
Based on Eqs.~\eqref{e:dumb_flow}, it is possible to obtain an analytic estimate for the mean tracer trajectory in the deterministic, far-field limit. To this end, we  consider initial conditions with $\hat{\bs n}=(1,0,0)$ as depicted in Fig.~\ref{fig01_Scatter} and approximate
\be\label{e:approx}
\bs r(t)\simeq \bs X(t)
\csp
\hat{\bs n} \hat{\bs r}\simeq \f{X(t)}{\sqrt{X(t)^2+Y_0^2}},
\ee
with $X(t)=X_0+tV$ denoting the swimmer's $x$-coordinate. The approximations~\eqref{e:approx} hold in the far-field scattering limit where  Eq.~\eqref{e:dumb_flow} is valid. Integrating the equations of motion~\eqref{e:LE-a} for $D_0=0$ yields the trajectory $\bs x(t)=(x(t),y(t),0)$ of the tracer particle in the $(z=0)$-plane
\bse\label{e:dumb_results}
\be
x(t)&\simeq&
-\gk_2A \left\{
\frac{A(2 X_0^2+Y_0^2)}{\left(X_0^2+Y_0^2\right)^{3/2}}-
\frac{A[2X(t)^2+Y_0^2]}{\left[X(t)^2+Y_0^2\right]^{3/2}}
\right\}, 
 \\
 y(t)&\simeq&
-\gk_2A \left\{\frac{AX_0 Y_0}{\left(X_0^2+Y_0^2\right)^{3/2}}-\frac{AX(t)Y_0}{\left[X(t)^2+Y_0^2\right]^{3/2}}\right\}.
\ee
\ese
Letting $t\to\infty$,  we see that asymptotically
\bse\label{e:dumb_asymptotics}
\be
x(t)&\to&
-\gk_2A \left[
\frac{A(2 X_0^2+Y_0^2)}{\left(X_0^2+Y_0^2\right)^{3/2}}
\right],
 \\
 y(t)&\to&
-\gk_2A \left[\frac{AX_0 Y_0}{\left(X_0^2+Y_0^2\right)^{3/2}}\right].
\ee
\ese
According to Eqs.~\eqref{e:dumb_asymptotics}, the tracer particle is displaced by a finite distance reflecting the fact that the effective interaction is short-range $\ga\ge 2$. Furthermore, Eq.~\eqref{e:dumb_asymptotics} predicts that, if the swimmer starts very far away, corresponding to the limit $X_0\to  -\infty$, the tracer returns to its initial condition $\bs x(0)=\bs 0$, thus forming a closed loop in the $(z=0)$-plane. 
\par
\begin{figure}[t!]
\centering
\includegraphics[width=8cm]{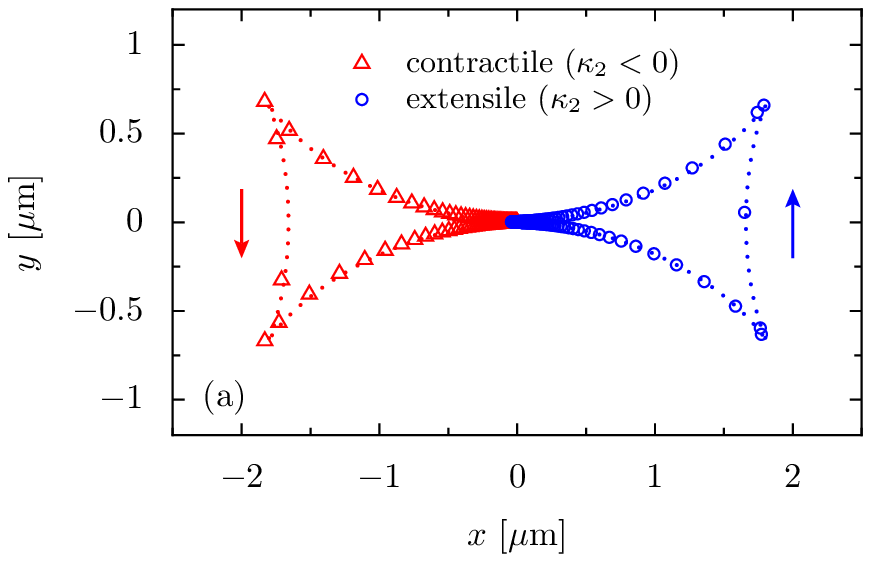}
\hspace{0.5cm}
\includegraphics[width=8cm]{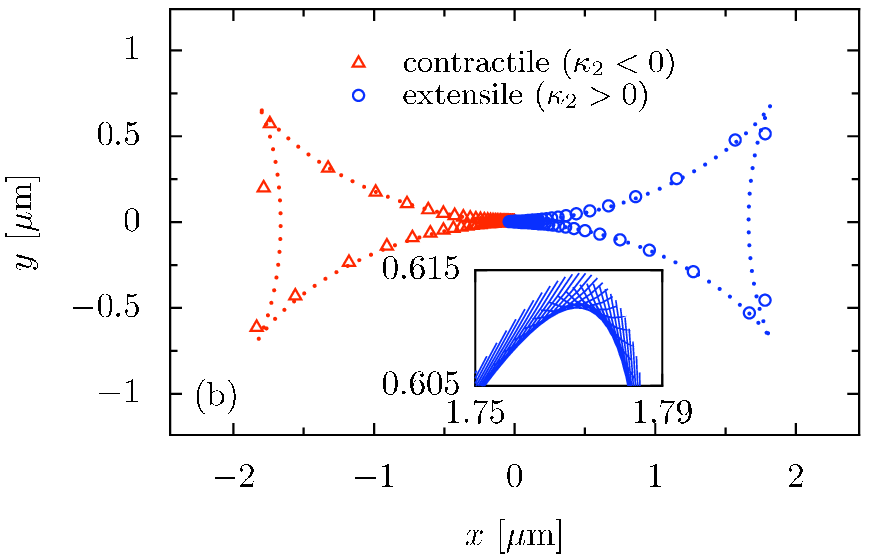}
\caption{
\label{fig0x_inch_long}
\emph{Deterministic tracer scattering by a dipolar 2-sphere swimmer.}
In the limit $D_0=0$ the tracer trajectories converge to closed loops if the force-free 2-sphere swimmer starts sufficiently far away from the tracer (arrows indicate how the loop is traversed).  The dotted lines in either diagram indicate the analytic estimate from Eq.~\eqref{e:dumb_results}. (a) Results for the stroke-averaged dynamics.   Symbols were obtained by numerically integrating Eq.~\eqref{e:LE-a} for $D_0=0$ for the dipolar flow field~\eqref{e:dumb_flow} using the same parameters as in Fig.~\ref{fig0x_inch_flow_averaged}. (b) Results for the time-resolved dynamics.  Symbols were obtained from numerical simulations of the tracer trajectories as described in Sec.~\ref{s:numerics}.  
The inset shows the details of the oscillatory motion of the tracer particle near  the turning point at the top-right corner of the extensile (right hand) path. The plots are based on initial conditions $X_0=-350A$, $Y_0=10A$ for the swimmer and $\bs x(0)=\bs 0$ for the tracer. The tracer trajectories are depicted for the time interval $[0,100\sec]$.  The swimmer parameters are the same as those in Fig.~\ref{fig0x_inch_flow_averaged}.  Tracer radius: $a_0=1\mu\met$. Simulation time step: $\Gd t=2.2\times 10^{-7}\sec$.  A comparison of the two diagrams implies that the stroke-averaged description is able to capture the main features of the mean particle trajectory.
}
\end{figure}

This is illustrated in Fig.~\ref{fig0x_inch_long}, which shows tracer trajectories in the $(x,y)$-plane. The dotted lines in both diagrams indicate the analytic estimates from  Eq.~\eqref{e:dumb_results}. Symbols in Fig.~\ref{fig0x_inch_long} (a) were obtained by numerically integrating Eq.~\eqref{e:LE-a} for $D_0=0$ for the time-averaged dipolar velocity field~\eqref{e:dumb_flow} using the same swimmer parameters as in Fig.~\ref{fig0x_inch_flow_averaged}. For comparison, we also plot in  Fig.~\ref{fig0x_inch_long} (b) the result from simulations that resolve the microscopic dynamics of the swimmer (details of the numerical implementation are given in Sec.~\ref{s:numerics}). In the latter case, one finds that the tracer particle performs a small oscillator motion around  its mean trajectory.   Hence,  by comparing Figs.~\ref{fig0x_inch_long} (a) and (b), one readily observes that the stroke-averaged description is able to capture the main features of the mean particle trajectory.

\begin{figure}[t!]
\centering
\includegraphics[width=8cm]{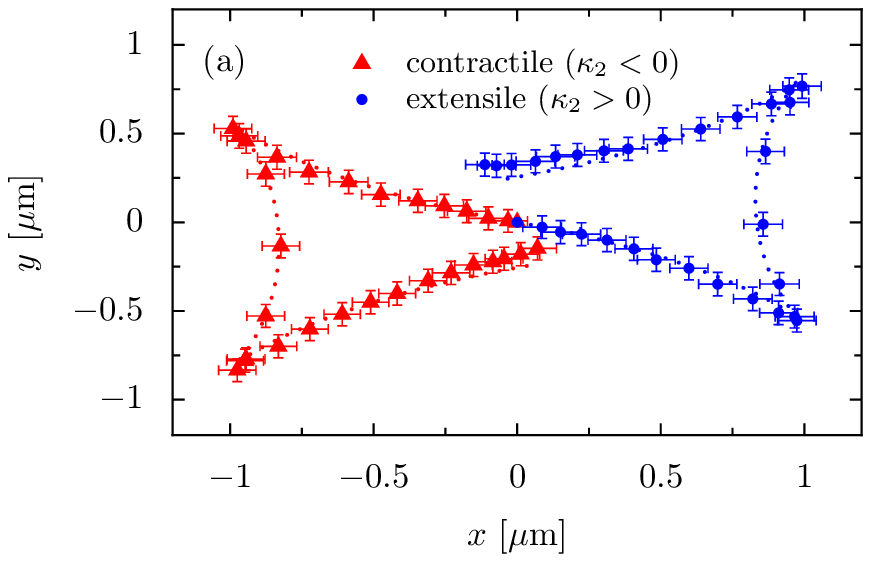}
\hspace{0.5cm}
\includegraphics[width=8cm]{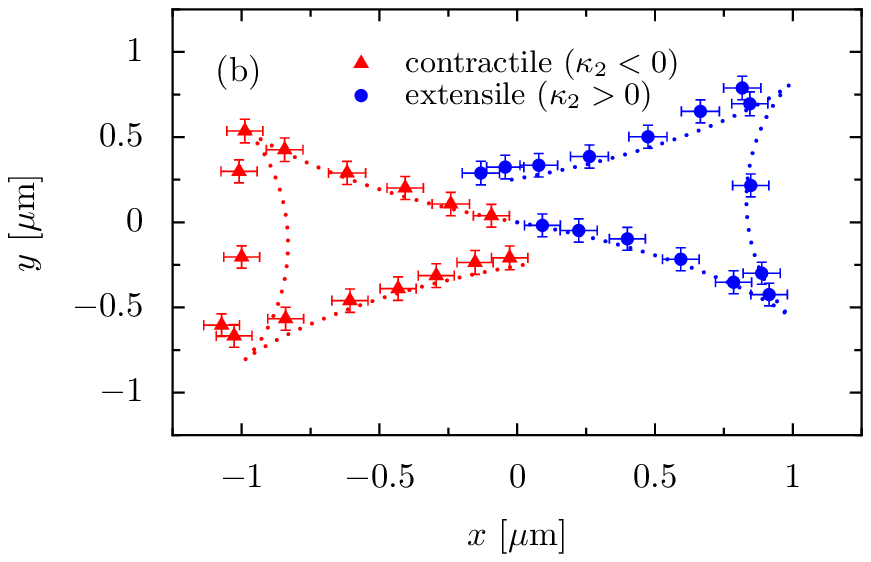}
\caption{
\label{fig0x_inch_noisy}
\emph{Brownian tracer scattering by a dipolar 2-sphere swimmer.} 
For $D_0>0$ loop-like scattering patterns can be recovered by averaging over many sample trajectories with identical initial conditions $\bs x(0)=\bs 0$. (a) Results for the stroke-averaged dynamics obtained by numerically integrating Eq.~\eqref{e:LE-a} with  $D_0=0.22\mu\met^2/\sec$.  (b) Results from the corresponding time-resolved simulations.   The filled symbols in both figures were obtained by averaging over 1000 sample trajectories and  dotted lines represent the analytic estimate from Eq.~\eqref{e:dumb_results}.  Statistical error bars correspond to the sample standard deviation, divided by $10\times \sqrt{t}$ for better visibility. Prior to rescaling the length of the error bars agreed with the theoretically  expected value~$\sqrt{2 D_0 t}$. Compared to Fig.~\ref{fig0x_inch_long}, we used the same swimmer parameters  but the swimmer was started nearer to the tracer at $X_0 =  -10A$ and sample trajectories were recorded for the shorter interval $t\in[0,10\sec]$. This explains why the loops are not fully closed.
}
\end{figure}

\paragraph*{Brownian motion effects.--}
As exemplified by Eq.~\eqref{e:dumb_flow}, the effective hydrodynamic flow fields  of self-swimming objects  decay more rapidly with distance than those of externally forced colloids. Hence, the hydrodynamic displacement of a tracer particle in the far-field flow of a force-free swimmer will, in general, be  smaller, so that thermal Brownian motion of the tracer becomes more  relevant. This raises the question as to whether loop-like structures could actually be observed under typical experimental conditions, e.g., similar to those considered by Leptos et al.~\cite{2009LeEtAl_Gold} who tracked the motions of algae  in water at room temperature. In order to estimate how Brownian effects may affect the loop-like structures, we numerically integrated the stochastic equations of motions~\eqref{e:LE-a} for both the time-averaged flow field~\eqref{e:dumb_flow} and the time-resolved swimmer dynamics, using a diffusion constant $D_0=0.22\mu\met^2/\sec$ corresponding to a tracer particle of radius $a_0=1\mu\met$ and water at room temperature.
\par
Figure~\ref{fig0x_inch_noisy}~(a) depicts the mean tracer trajectories  obtained numerically for the time-averaged model for an extensile (red triangles) and a contractile swimmer (blue circles). Figure~\ref{fig0x_inch_noisy}~(b) shows the corresponding results for the time-resolved swimmer dynamics. The numerical data points (filled symbols) in Figs.~\ref{fig0x_inch_noisy} were obtained by averaging over 1000 sample  trajectories with identical initial conditions. In both cases, we used the same swimmer parameters as in Figs.~\ref {fig0x_inch_long}, but the swimmer was started closer to the tracer at $X_0=-35A$. The simulation time per run was limited to $10\sec$. For comparison, isolated Chlamydomonas algae typically swim on an almost straight line for about $5-10\sec$~\cite{2009PoEtAl_Gold} before changing their directions due to tumbling and rotational Brownian motion.  The statistical error bars in Figs.~\ref{fig0x_inch_noisy} correspond to the sample standard deviation and were divided by $10\times \sqrt{t}$ for better visibility. Prior to this rescaling the length of the error bars agrees well with the theoretically expected value~$\sqrt{2 D_0 t}$. 
\par
To summarize, the diagrams in Fig.~\ref{fig0x_inch_noisy}  suggest that after averaging over a sufficiently large sample size the loop-like patterns can be recovered. Individual trajectories may however look very different due to Brownian motion. 

\subsection{(Quasi-)quadrupolar three-sphere swimmer}
\label{s:quadrupolar}
\begin{figure}[t!]
\centering
\includegraphics[width=5.8cm]{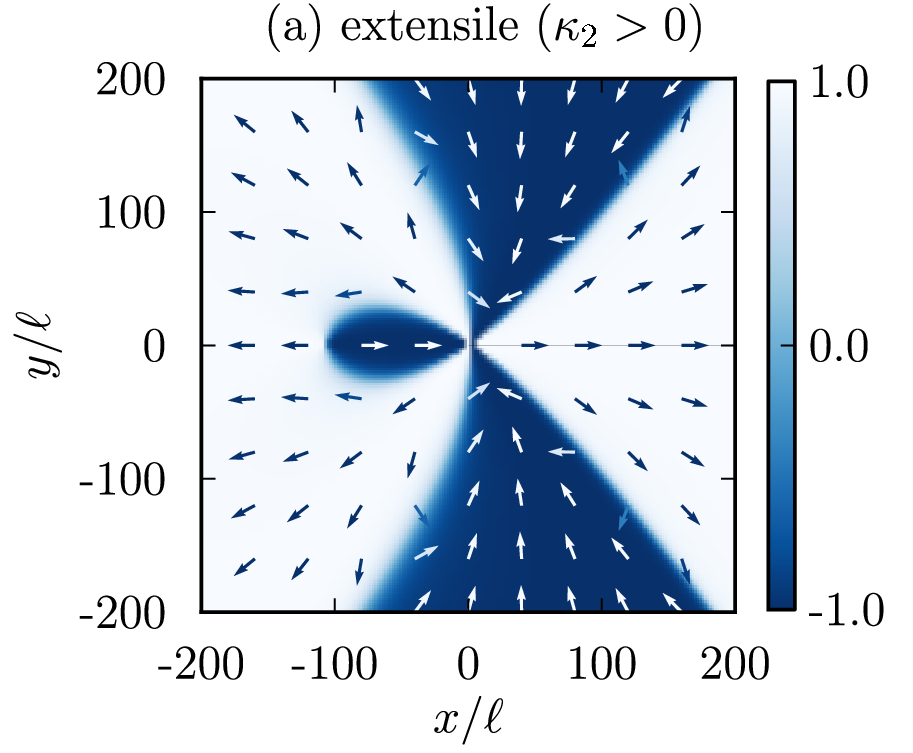}
\includegraphics[width=5.8cm]{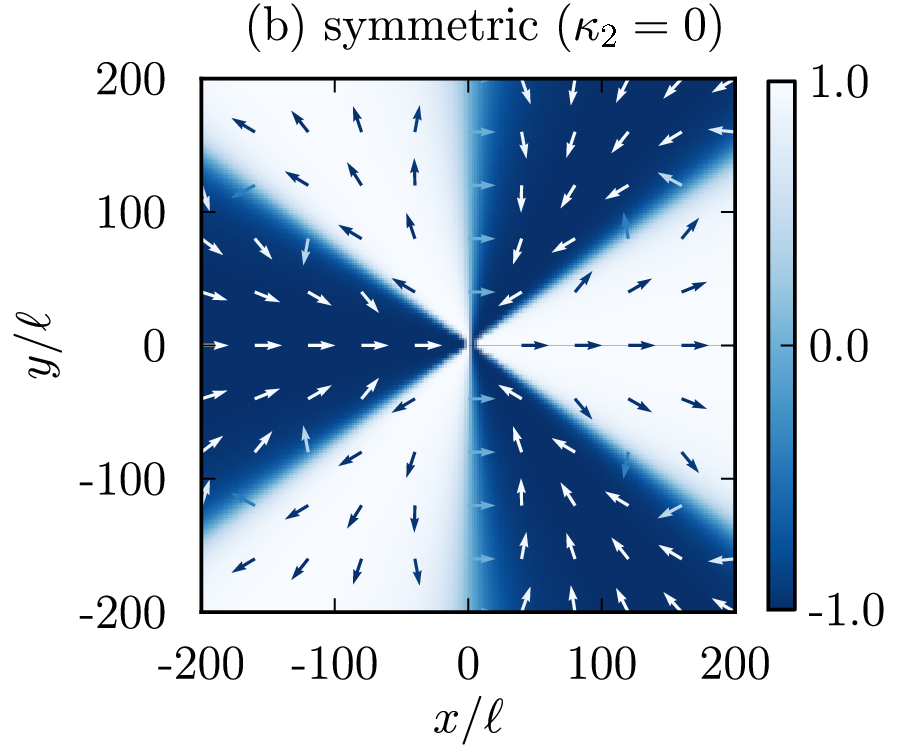}
\includegraphics[width=5.8cm]{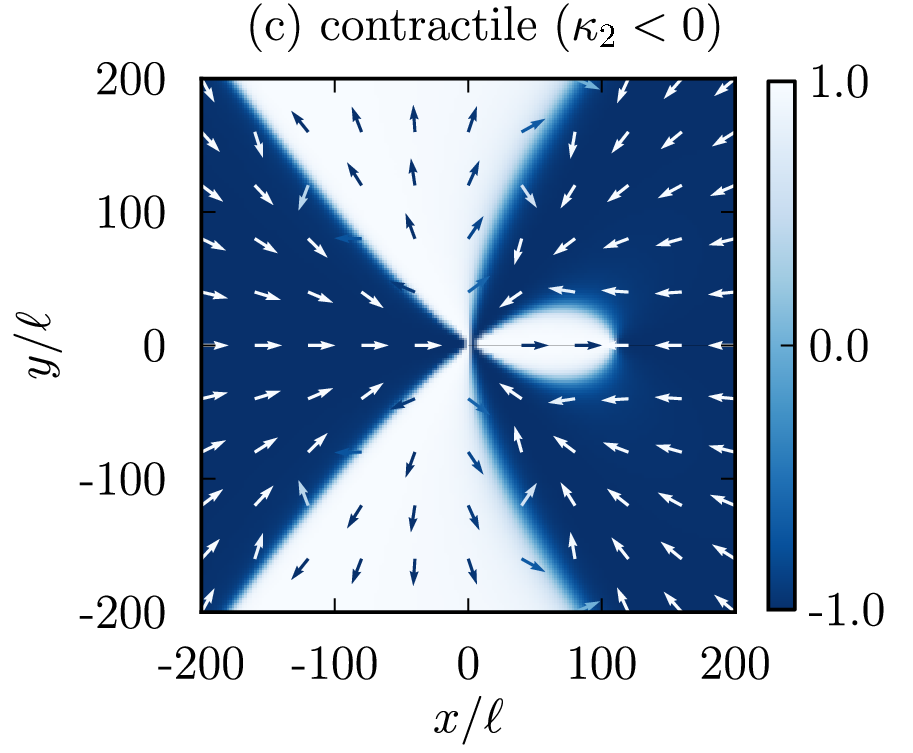}
\caption{
\label{fig0x_gol_flow}
\emph{Quasi-quadrupolar fluid flows generated by 3-sphere swimmers.}
Structure of the stroke-averaged flow field~\eqref{e:gol_flow}  for three different parameter sets: (a)~extensile pusher with $0.2\ell=\xi_R>\xi_F=0.1\ell$, (b)~symmetric swimmer  with $\xi_R=\xi_F=\sqrt{0.02}\ell$, and (c)~contractile puller  with $0.1\ell=\xi_R<\xi_F=0.2\ell$.  In all three cases, the swimmer is located at the origin, points along the $x$-axis, and swims in the positive $x$-direction. The flow fields are rotationally invariant about the $x$-axis. Arrows indicate the local direction $\hat{\bs v}:=\bs v/|\bs v|$ of the stroke-averaged flow field~\eqref{e:gol_flow}. Dark (bright) areas show the projection  $(\hat{\bs r}\hat{\bs v})$,  corresponding to  regions where the mean flow points towards (away from) the swimmer. Note that in all three cases the quadrupolar contribution dominates at intermediate distances. Swimmer parameters: $\ell=10\mu\met, A=0.2\ell, \go=10^3\sec^{-1}, \Gd\phi=\pi/2$, corresponding to $V\simeq 12\mu\met/\sec$. 
}
\end{figure}

As the final example, we consider the scattering of a tracer by a linear  3-sphere swimmer~\cite{2008GoAj,2009AlPoYe}, as illustrated in Fig.~\ref{fig03_swimmer} (b). The main difference when compared to the 2-sphere swimmer occurs because the flow field of this swimmer is essentially quadrupolar,  see Fig.~\ref{fig0x_gol_flow}.  The lengths of  the rear leg  ($R$) and the front  leg ($F$) of the 3-sphere swimmer vary in time as
\be\label{e:3-leg}
L_{F,R}(t)=\ell + \xi_{F,R} \sin(\go t +\phi_{F,R})
\csp
\ell\gg\xi_{R,F}>0,
\quad 
\ell\gg A>0,
\ee
where $\ell$ is the mean leg length and $\xi_{R,F}$ denotes the oscillation amplitude. If the initial phases $\phi_{R,F}$  are chosen such that
\be
\Gd\phi:=\phi_R-\phi_F>0,
\ee
then the swimmer moves with stroke-averaged speed 
\be\label{e:3-velo}
V=\f{7}{24} A\go \sin(\Gd\phi) \left(\f{\xi_F\xi_R}{\ell^2}\right)
\ee
in the direction of the front leg. The time averaged flow field at distances far away from the 3-sphere swimmer is~\cite{2009AlPoYe}
\be
\bs v(\bs x(t)| \bs X(t),\bs V)
&=&
V\gk_2 \left(\f{ A}{|\bs r(t)|}\right)^2
\left[   
3 (\hat{\bs n} \hat{\bs r})^2-1     
\right]  \hat{\bs r}
+
\notag\\
&&
V \gk_3 \left(\f{A}{|\bs r(t)|}\right)^3
\left\{
\left[ 15 (\hat{\bs n} \hat{\bs r})^3-9  (\hat{\bs n} \hat{\bs r})   \right]\hat{\bs r}
-
\left[  3 (\hat{\bs n} \hat{\bs r})^2-1     
\right]  {\hat{\bs n}}
\right\}
\label{e:gol_flow}
\ee
with unit vectors $\hat{\bs n}$ and $\hat{\bs r}$ as defined in Eq.~\eqref{e:unit_vectors}. The constants $\gk_{2,3}$ are determined by the swimmer's geometry and read
\be
\gk_2=\f{9}{32} \f{\xi_R^2-\xi_F^2}{A\ell}
\csp
\gk_3=\f{51}{56}\left( \f{\ell}{A}\right)^2.
\ee
The quadrupole coefficient is always positive, $\gk_3>0$, but we can have either $\gk_2>0$ (extensile \lq pusher\rq), $\gk_2=0$ (symmetric quadrupolar swimmer) or $\gk_2<0$ (contractile \lq puller\rq), depending on the choice of the oscillation amplitudes $\xi_{R,F}$. It is important  to note, however, that the  geometric restriction $\ell>2A+\xi_R+\xi_F$ implies that  $\gk_2\ll \gk_3$. Consequently,  the \emph{quadrupolar} $\gk_3$-term effectively dominates the flow field~\eqref{e:gol_flow} unless one considers unreasonably large distances~\cite{2009AlPoYe}. This is illustrated in Fig.~\ref{fig0x_gol_flow} which shows the normalized flow field $\hat{\bs v}:=\bs v/|\bs v|$ from Eq.~\eqref{e:gol_flow} and also its projection onto the unit vector $\hat{\bs r}(t)$ for $\gk_2>0$, $\gk_2=0$, and $\gk_2<0$, respectively. 
 
\begin{figure}[t]
\centering
\includegraphics[width=8cm]{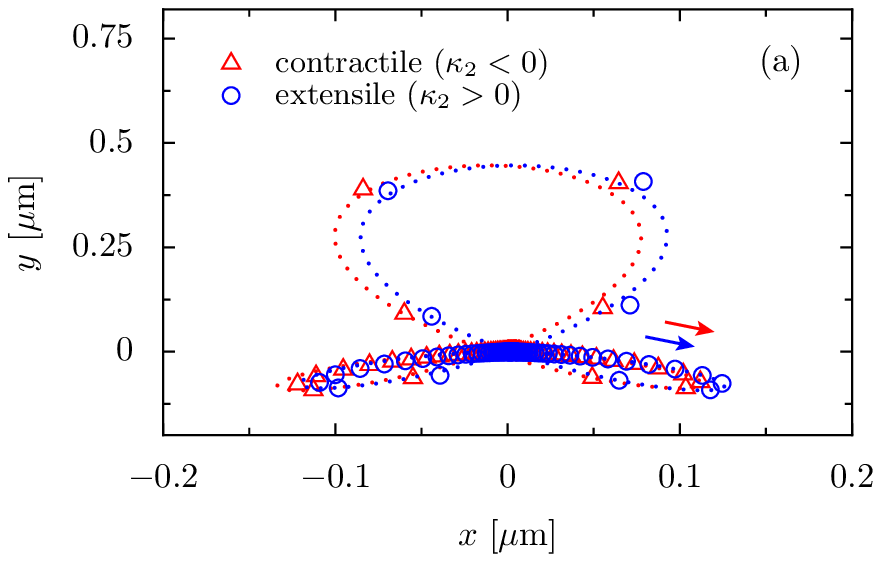}
\hspace{0.5cm}
\includegraphics[width=8cm]{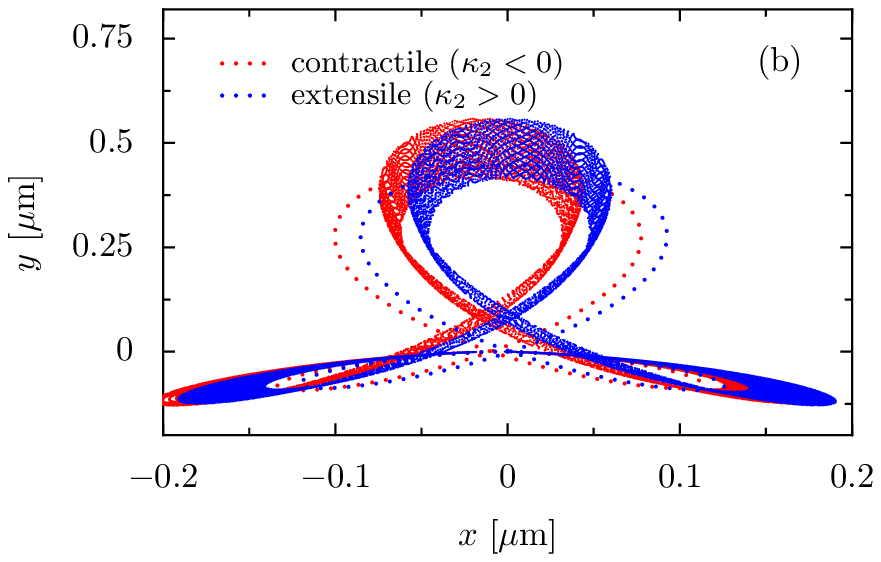}
\caption{
\label{fig0x_gol_long} 
\emph{Deterministic tracer scattering by a (quasi-)quadrupolar 3-sphere swimmer.}
In the deterministic limit $D_0=0$,  the tracer trajectories converge to closed loops if the force-free 3-sphere swimmer starts sufficiently far away from the tracer, $X_0\to-\infty$.  The dotted lines in either diagram represent the analytical estimate from Eq.~\eqref{e:gol_results}. (a) Results from the stroke-averaged dynamics for the time interval $[0,55\sec]$. The tracer particle starts at $\bs x(0)=\bs 0$ and the swimmer at $X_0=-150A$, $Y_0=10A$. Symbols were obtained by numerically integrating Eq.~\eqref{e:LE-a} for $D_0=0$ using the averaged flow field~\eqref{e:gol_flow}.  (b)~Results from the corresponding time-resolved simulation for the same time interval and same initial conditions for  the geometric center of the swimmer. The tracer oscillates due to the periodicity of the swimming stroke.  Swimmer parameters: $\ell=10\mu\met, A=0.2\ell, \go=10^3\sec^{-1}, \Gd\phi=\pi/2$, which results in $V\simeq 12\mu\met/\sec$, $\gk_2\simeq \pm 0.04$, $\gk_3\simeq 22$.   Tracer radius: $a_0=1\mu\met$. Simulation time step: $\Gd t=2.2\times 10^{-7}\sec$.
}
\end{figure}

 \paragraph*{Tracer motion in the deterministic limit.--}
Based on Eq.~\eqref{e:gol_flow} it is again possible to derive an analytic estimate for the mean tracer trajectory. To this end, we  consider initial conditions with $\hat{\bs n}=(1,0,0)$ as depicted in Fig.~\ref{fig01_Scatter} and make use of the approximation~\eqref{e:approx}, which holds in the far-field scattering limit where  Eq.~\eqref{e:gol_flow} is valid. Integrating the equations of motion~\eqref{e:LE-a} for $D_0=0$, we find that the trajectory of the tracer particle in the $(z=0)$-plane is given by $\bs x(t)=(x(t),y(t),0)$, where
\bse\label{e:gol_results}
\be
x(t)&\simeq&
-\gk_2A \left\{
\frac{A(2 X_0^2+Y_0^2)}{\left(X_0^2+Y_0^2\right)^{3/2}}-
\frac{A[2X(t)^2+Y_0^2]}{\left[X(t)^2+Y_0^2\right]^{3/2}}
\right\} 
 +
 \gk_3 A\left\{
 \frac{A^2X_0\left(2 X_0^2-Y_0^2\right)}{\left(X_0^2+Y_0^2\right)^{5/2}}-
 \frac{A^2 X(t) \left[2 X(t)^2-Y_0^2\right]}{\left[X(t)^2+Y_0^2\right]^{5/2}}
 \right\},\quad
 \\
 y(t)&\simeq&
-\gk_2A \left\{\frac{AX_0 Y_0}{\left(X_0^2+Y_0^2\right)^{3/2}}-\frac{AX(t)Y_0}{\left[X(t)^2+Y_0^2\right]^{3/2}}\right\}
+
\gk_3 A
 \left\{\frac{A^2(2 X_0^2-Y_0^2)Y_0}{\left(X_0^2+Y_0^2\right)^{5/2}}-
 \frac{A^2[2X(t)^2-Y_0^2]Y_0}{\left[X(t)^2+Y_0^2\right]^{5/2}} \right\}.
\ee
\ese
Note that the dipolar $\gk_2$-contributions are formally equivalent to the result~\eqref{e:dumb_results} for the 2-sphere swimmer, but with $V$ and $\gk_2$ now determined by the geometry of the 3-sphere swimmer. Letting $t\to\infty$ we find that asymptotically
\bse\label{e:gol_asymptotics}
\be
x(t)&\to&
-\gk_2A \left[
\frac{A(2 X_0^2+Y_0^2)}{\left(X_0^2+Y_0^2\right)^{3/2}}
\right]
 +
 \gk_3 A\left[
 \frac{A^2X_0\left(2 X_0^2-Y_0^2\right)}{\left(X_0^2+Y_0^2\right)^{5/2}}
 \right] ,
 \\
 y(t)&\to&
-\gk_2A \left[\frac{AX_0 Y_0}{\left(X_0^2+Y_0^2\right)^{3/2}}\right]
+
\gk_3 A
 \left[\frac{A^2(2 X_0^2-Y_0^2)Y_0}{\left(X_0^2+Y_0^2\right)^{5/2}} \right ],
\ee
\ese
which implies that the tracer particle is displaced by a finite distance  due to the effective short-range interaction with $\ga\ge 2$.  Equations~\eqref{e:gol_asymptotics} predict that, if the swimmer starts very far away, corresponding to the limit $X_0\to  -\infty$, the tracer returns to its initial condition $\bs x(0)=\bs 0$, thus forming a closed loop in the $(z=0)$-plane. This is  similar to our previous result~\eqref{e:dumb_asymptotics} for the 2-sphere dipolar swimmer. However,  comparing with Fig.~\ref{fig0x_inch_long} (a), the shape of the loop in Fig.~\ref{fig0x_gol_long} (a) differs significantly,  because the flow field~\eqref{e:gol_flow}  of the 3-sphere swimmer is dominated by the quadrupolar $\gk_3$-contribution (for the initial conditions considered).
\par

The analytical estimate~\eqref{e:gol_results} is based on the stroke-averaged field and it is interesting to compare with simulations for the corresponding time-resolved swimmer. The results for the latter case are depicted in Fig.~\ref{fig0x_gol_long} (b). Due to the oscillations of the swimmer's legs, the tracer particle performs an oscillatory motion around its mean trajectory. The relative magnitude of the oscillations is bigger than in the dipolar 2-sphere swimmer case, since the effective quadrupolar flow field of the 3-sphere swimmer decays more rapidly, resulting in a smaller absolute loop-size at similar initial conditions and swimmer speeds.  As evident from  Figs.~\ref{fig0x_gol_long}, the stroke-averaged result~\eqref{e:gol_results} captures the main features of the mean tracer motion in the time-resolved flow field but  there is a quantitative difference  of about 20 to 30 percent. However, we observe that, asymptotically, the time-resolved tracer trajectory  approximately returns to its initial position, if the swimmer starts sufficiently far away from the tracer particle.

\begin{figure}[t]
\centering
\includegraphics[width=8cm]{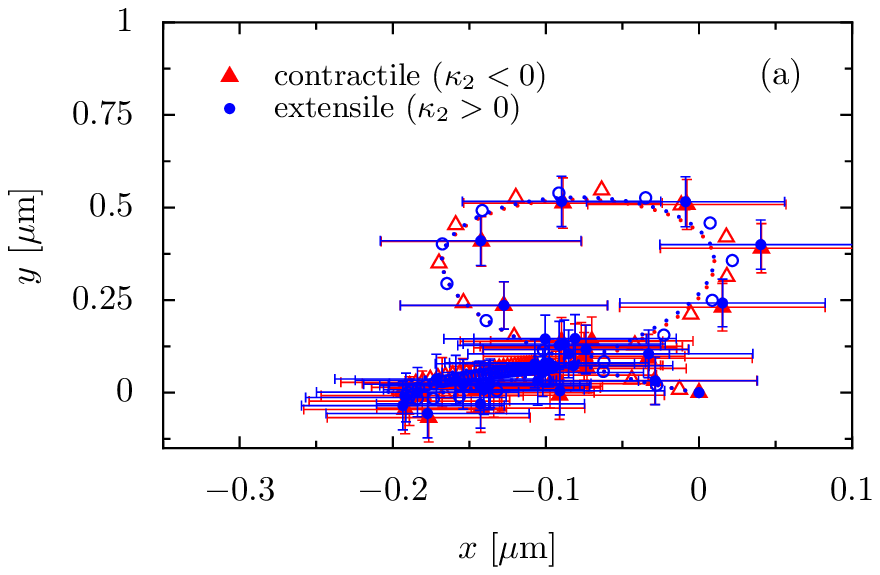}
\hspace{0.5cm}
\includegraphics[width=8cm]{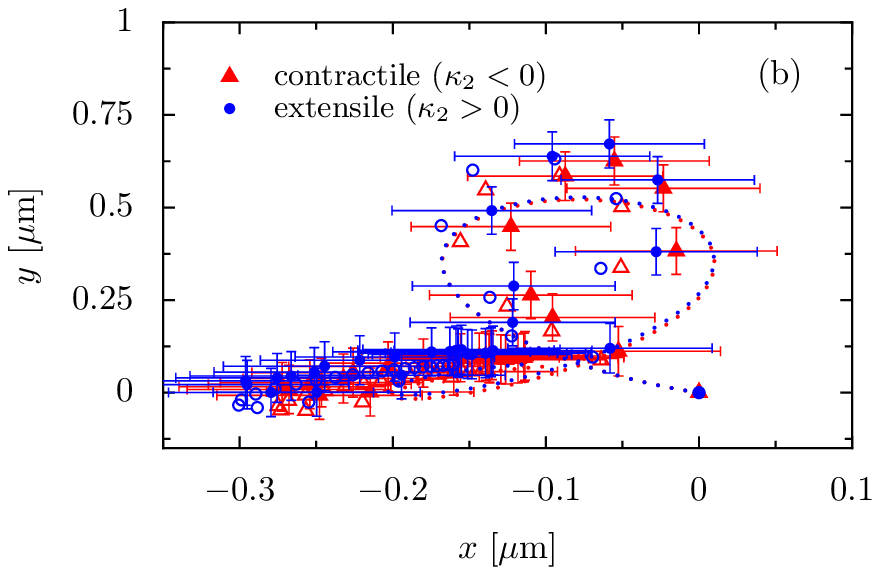}
\caption{
\label{fig0x_gol_noisy}
\emph{Brownian tracer scattering by a (quasi-)quadrupolar 3-sphere swimmer.}
For $D_0>0$ loop-like scattering patterns can be recovered by averaging over many sample trajectories with identical initial conditions $\bs x(0)=\bs 0$. (a) Results for the stroke-averaged dynamics obtained by numerically integrating Eq.~\eqref{e:LE-a} with  $D_0=0.22\mu\met^2/\sec$.  (b) Results from the corresponding time-resolved simulations.   The filled symbols in both figures were obtained by averaging over 1000 sample trajectories, unfilled symbols show the results of simulations with $D_0=0$, and  dotted lines represent the analytic estimate from Eq.~\eqref{e:gol_results}.  Statistical error bars correspond to the sample standard deviation, divided by $10\times \sqrt{t}$ for better visibility. Prior to rescaling the length of the error bars agreed with the theoretically  expected value~$\sqrt{2 D_0 t}$. Compared to Fig.~\ref{fig0x_gol_long}, we used the same swimmer parameters  but the swimmer was started nearer to the tracer at $X_0 =  -10A$ and sample trajectories were recorded for the shorter interval $t\in[0,11\sec]$. This explains why the loops are not fully closed.
}
\end{figure}

\paragraph*{Brownian motion effects.--}
When Brownian motion is taken into account, corresponding to $D_0>0$ in Eq.~\eqref{e:LE-a}, individual tracer trajectories may differ strongly from each other. As for the 2-sphere dipolar swimmer, loop-shaped scattering patterns can be reconstructed by averaging over a tracer ensemble  with identical initial conditions. Figures~\ref{fig0x_gol_noisy} (a) and (b)  show mean tracer trajectories obtained by averaging over 1000 sample  trajectories. We used the same swimmer parameters as in Figs.~\ref {fig0x_gol_long}, but the swimmer was started closer to the tracer at $X_0=-10A$. The simulation time per run was limited to $11\sec$, as this corresponds roughly to the time scale  during which an isolated alga performs a quasilinear motion~\cite{2009PoEtAl_Gold}.  The statistical error bars in Figs.~\ref{fig0x_gol_noisy} illustrate the sample standard deviation and were divided by $10\times \sqrt{t}$ for better visibility. Prior to this rescaling the length of the error bars agreed well with the value~$\sqrt{2 D_0 t}$ expected theoretically.  In both Fig.~\ref{fig0x_gol_noisy}~(a) and~(b), we notice a small, systematic drift of the mean Brownian trajectories compared to the corresponding deterministic trajectories (unfilled symbols) and the analytic estimate (dotted lines). The shift occurs in the direction of the swimmer motion and is probably caused by a noise-induced Stokes drift~\cite{1998LyJa} and an additional, noise-induced bias due to the gradients of the flow field. 

\section{Numerical methods}
\label{s:numerics}
One of our objectives was to compare the tracer dynamics in the stroke-averaged and explicitly time-dependent flow fields of dipolar and quadrupolar model swimmers. For this purpose, we focussed on multi-sphere swimmer models that can be implemented numerically in Brownian dynamics simulations~\cite{2009DuZa,2010PuDuYe,2009TeZiLo,1988BrBo}.
\par 
The time-resolved, \lq microscopic\rq\space dynamics of the 2-sphere swimmer was computed numerically by using a spring-based model~\cite{2010PuDuYe}. Within this approach, the two spheres are linked by  a harmonic spring  (stiffness $k_0$) with periodically varying equilibrium length $L(t)$ determined by Eq.~\eqref{e:2-length}. Compared to the implementation in Ref.~\cite{2010PuDuYe}, the only technical difference  arises because our dipolar 2-sphere swimmer is characterized by an oscillating sphere radius, leading to time-dependent Stokes friction coefficients  $\gc_i(t)= 6\pi\eta A_i(t)$, $i=1,2$. If the stiffness parameter $k_0$ of the spring model is chosen large enough, 
the swimmer essentially behaves like a shape-driven swimmer (in our simulations we fixed $k_0= 0.001\, \kg/\sec^2$).
\par
The time-resolved microscopic dynamics of the linear 3-sphere swimmer can be  implemented in a similar manner~\cite{2009DuZa}.  Analogous to the 2-sphere swimmer case, two neighboring spheres (constant radius $A$) are connected by  a time-dependent harmonic spring (the \lq leg\rq) with the same value of the stiffness parameter  $k_0$ as above.  Each spring  has  an oscillating equilibrium length given by Eq.~\eqref{e:3-leg}.   The velocity field acting on the tracer  and the hydrodynamic interactions between the spheres were calculated using the Rotne-Prager-Yamakawa-Mazur tensor~\cite{1969RoPr,1970Ya,Oseen,HappelBrenner,1982Ma}. 
\par
Upon adopting the same swimmer parameters for both the microscopic and the averaged model, one finds that the average swimmer speed measured in the microscopic simulations agrees well with the analytical estimates from Eqs.~\eqref{e:dumb_vel} and \eqref{e:3-velo}, respectively. We also compared the Lagrangian net fluid flow, which was obtained from the time-resolved simulations by tracking a grid of non-interacting tracer particles over a stroke period $[t,t+2\pi/\go]$. The dipolar and quadrupolar flow field structures obtained by this method were found to be identical to those from Eqs.~\eqref{e:dumb_flow} and~\eqref{e:gol_flow} depicted in Figs.~\ref{fig0x_inch_flow_averaged} and~\ref{fig0x_gol_flow}.
\par
The equations of motion for both swimmer and tracer  were numerically integrated with a two-step Heun integrator  using Nvidia's CUDA toolkit. The simulations for the time-resolved model  were performed on a consumer-level GTX 275 GPU embedded in an Intel i860 PC running Gentoo Linux. The GPU-based parallel computation of sample trajectories in the case of Brownian tracer particles leads to a considerable speed-up (up to a factor of a few hundreds) compared with a conventional un-parallelized CPU-based implementation~\cite{2009JaKo}.

\section{Summary \& discussion}
We have studied the hydrodynamic scattering of a tracer particle by  different types of model swimmers.
  We showed, both analytically and numerically, that scattering by a force-free swimmer may lead to quasi-closed, loop-shaped tracer trajectories. The shapes and orientations of the loops are a signature of the properties of the stroke-averaged flow field.  A detailed comparison with the time-resolved scattering dynamics implies that a time-averaged description is sufficient to capture the main features of swimmer-tracer scattering. Our analysis has focussed on two specific 2-sphere and 3-sphere swimmer models that are analytically tractable and can be implemented numerically. However,  the result~\eqref{e:gol_results} is generic;  it holds for the large class of low Reynolds number swimmers that generate a dipolar or quadrupolar flow as given in Eq.~\eqref{e:gol_flow}.
\par
We have also compared the scattering due to swimmers to that resulting when a colloid is pulled past a tracer. This allowed us to demonstrate the role played by the range of the interactions between the scatterer and the tracer. For interactions decaying as  $\sim |\bs r|^{-\alpha}$, the tracer is slowly dragged to infinity by the particle for the colloidal case $\alpha=1$ (within the zero Reynolds number approximation), whereas its displacement remains finite for scattering by a force-free swimmer with $\alpha \ge 2$.
\par
We conclude the paper by addressing potential experimental realizations and by commenting on generalizations. The 3-sphere swimmer considered in the last part of the paper  could possibly be realized in practice by adapting the experimental setup of Leoni et al.~\cite{2009LeEtAl}, who constructed a colloidal linear 3-sphere pump using optical tweezers, to mimic a self-motile, force-free swimmer. With regard to hydrodynamic scattering experiments, an advantage of artificially created  colloidal microswimmers lies in the fact that one could tune the fluid viscosity to suppress Brownian motion effects in the tracer dynamics. In the case of a  colloidal 3-sphere swimmer one should expect to observe loops that are shaped similar to those in Fig.~\ref{fig0x_gol_long}.
\par
In general, Brownian motion can obscure individual tracer trajectories, but our analysis shows that  averaging over many trajectories may allow one to reconstruct the loops even from noisy data. This suggests that such patterns could also  be studied in biophysical experiments similar to those of Leptos et al.~\cite{2009LeEtAl_Gold}. These authors investigated  how  the diffusion of (an ensemble of)  tracer particles in water is affected by the presence of  self-motile unicellular~\emph{Chlamydomonas reinhardtii} algae.   Indeed loop-shaped tracer  trajectories (or, at least, segments thereof) have already been observed in their experiments, see inset of Fig. 1(c) in Ref.~\cite{2009LeEtAl_Gold}.   By systematically analyzing the trajectories of individual tracer particles in a sufficiently large, dilute suspension of algae, one could achieve the required sample size to reconstruct the exact structure of the loops and thus infer details about the fluid flow generated by the microorganisms. 
 \par
 Lastly it is plausible to assume that an improved understanding of the underlying, elementary swimmer-tracer scattering  processes may eventually help to model quantitatively  the anomalous diffusive behavior of tracer particles in dilute, active suspensions~\cite{2009LeEtAl_Gold}. The aim would be to derive an analytic expression for  the time-dependent position probability density function of  a tracer particle by averaging a representative trajectory, resulting from a  superposition of the flow field contributions from many algae, over a set of suitably chosen random initial configurations.  However, the averaging procedure is both mathematically and computationally demanding; we hope to tackle this problem in  future work.

\paragraph*{Acknowledgements.--}
This project was supported by the ONR, USA (J.~D.) and,  in part, by the Natural Sciences and Engineering Research Council of Canada (I.~M.~Z.). V.~P. acknowledges support from the
United States Air Force Institute of Technology.   The views expressed
in this paper are those of the authors and do not reflect the official
policy or position of the United States Air Force, Department of
Defense, or the U.S. Government.


\providecommand*{\mcitethebibliography}{\thebibliography}
\csname @ifundefined\endcsname{endmcitethebibliography}
{\let\endmcitethebibliography\endthebibliography}{}

\end{document}